\documentclass[twocolumn]{aastex631}

\usepackage{hyperref,amsmath,amssymb}
\usepackage{graphicx}

\numberwithin{equation}{section}

\begin{document}

\title{Chandra X-ray Observations of the Pulsar Wind Nebula within CTA 1 }

\author{Seth Gagnon}
\affiliation{Department of Physics, The George Washington University, 725 21st St, NW, Washington, DC 20052, USA}

\author{Oleg Kargaltsev}
\affiliation{Department of Physics, The George Washington University, 725 21st St, NW, Washington, DC 20052, USA}

\author{Jason Alford}
\affiliation{Division of Science, New York University Abu Dhabi, PO Box 129188, Abu Dhabi, UAE} 
\affiliation{Center for Astrophysics and Space Science (CASS), New York University Abu Dhabi, PO Box 129188, Abu Dhabi, UAE}

\author{Joseph Gelfand}
\affiliation{Division of Science, New York University Abu Dhabi, PO Box 129188, Abu Dhabi, UAE}
\affiliation{Center for Astrophysics and Space Science (CASS), New York University Abu Dhabi, PO Box 129188, Abu Dhabi, UAE}

\author{Alexander Lange}
\affiliation{Department of Physics, The George Washington University, 725 21st St, NW, Washington, DC 20052, USA}

\shorttitle{Chandra Observations of CTA 1}
\shortauthors{Gagnon et al.}

\begin{abstract}

We present deep \textit{Chandra}
observations of the pulsar wind nebula (PWN) powered by PSR J0007+7303 in the composite supernova remnant CTA~1. 
The merged ACIS image 
shows a $\sim20''$ jet extending south of the pulsar and bending toward the southwest, a faint counter-jet to the north, and a compact torus oriented approximately perpendicular to the jet axis. Using an archival observation from 2003 we perform  relative astrometry over a $\sim20$ yr baseline and constrain the pulsar's transverse velocity to $\lesssim 200~\mathrm{km~s^{-1}}$ at the distance of 1.4 kpc at 95\% confidence. 
Spatially resolved spectroscopy 
shows hard spectra for the jet and torus (photon indicies $\Gamma \approx 1.2$--1.4) and a softer spectrum for the extended nebula ($\Gamma = 1.85 \pm 0.11$), 
indicating minimal radiative cooling in the compact regions. Modeling of the torus, associated with the termination shock, 
as an inclined circle yields a viewing angle $\zeta \approx 50^\circ$.
The outer gap and two-pole caustic pulsar emission 
models then imply 
a moderate magnetic inclination ($\alpha \sim 20^\circ$--$70^\circ$). Broadband spectral energy distribution (SED) modeling from radio to PeV $\gamma$-rays 
for a one-zone leptonic scenario yields a low magnetic field ($B \approx 1.4$--$3.2~\mu\mathrm{G}$) and a high electron cutoff energy ($E_{\rm cut} \sim 0.2$--$0.3~\mathrm{PeV}$), indicating that the magnetic field decreases rapidly outside of the compact nebula.
These results establish CTA~1 as a young, low X-ray efficiency PWN with a hard injection spectrum capable of accelerating particles to PeV energies.

\end{abstract}


\section{Introduction} \label{sec:intro}

Pulsars are among nature’s most powerful particle accelerators, capable of producing particles with energies up to a
few PeV. As a pulsar spins down, most of its rotational energy
is converted into a magnetized ultrarelativistic-particle wind,
whose synchrotron emission can be seen from radio to $\gamma$-rays
as a pulsar wind nebula (PWN). To date, the largest number of
PWNe have been discovered in X-rays with Chandra X-ray
Observatory (CXO hereafter; see
\citealt{2008AIPC..983..171K,2017JPhCS.932a2050K,2017SSRv..207..175R}
for reviews),
thanks to its unrivaled angular resolution and the very low
ACIS background.
The remarkably sharp images of pulsar wind nebulae 
\citealt{2017SSRv..207..175R}
from CXO allow us to learn
about the pulsar wind properties and about the intrinsic properties of the pulsar \citep{2015SSRv..191..391K,
2017JPlPh..83e6301K}. For example,
a PWN's morphology, surface brightness, and spectrum can constrain the magnetic field strength, electron
energies, the pulsar velocity imparted via a kick during the SN explosion, and the orientations of
the pulsar's spin/magnetic axes. 
Models of pulsar magnetospheres 
use the viewing angle  $\zeta$ (between the spin
axis and the line-of-sight)
and 
the magnetic inclination angle $\alpha$ (between
the magnetic and spin axes)  
to calculate the radio and $\gamma$-ray pulse profiles
(e.g., 
\citealt{Watters_2009}, 
\citealt{2015A&A...575A...3P}, \citealt{2016MNRAS.457.2401C}, 
\citealt{2023ApJ...954..204K},
\citealt{2024A&A...687A.169P}). However, these models have additional fitting parameters and use different assumptions
about the magnetosphere geometry and the location of the emission region (see 
\citealt{2016JPlPh..82c6306H},
\citealt{2022ARA&A..60..495P},
\citealt{2024arXiv240306180C}
for reviews). Therefore, obtaining
independent constraints on $\zeta$ and $\alpha$ from PWN morphologies in high-resolution images provides important independent diagnostics for competing pulsar magnetosphere models.
Modeling of CXO images of PWNe 
(\citealt{2004ApJ...601..479N}, \citealt{2008ApJ...673..411N})
have
demonstrated that faint compact PWN structures (tori and jets) can be used to infer the
3D orientation of the pulsar spin axis.
Such fitting
also provides an accurate measurement of the termination shock scale for the equatorial wind and
relative luminosities of the tori and jet components (related to the angle $\alpha$; 
\citealt{2016MNRAS.462.2762B}). For example, the
relative strengths of the jets and tori can vary: some PWNe appear to be dominated by equatorial
outflows (e.g., the Crab, Vela, 3C58, G54.1+0.3), while others are dominated by jets (e.g., B1509–58, Kes 75, G11.2–0.3). Also, some PWNe exhibit a single torus (e.g., the Crab; \citealt{2000ApJ...536L..81W}) while others
show double tori (e.g., Vela and the “Dragonfly Nebula”; 
\citealt{2003ApJ...591.1157P} and \citealt{2008ApJ...680.1417V}, respectively). 
PWN morphology can also be affected by supersonic pulsar motion. If such a PWN features
prominent jets, they become swept back by the ram pressure of the oncoming medium.
If the tangential velocity component is measured via the proper motion, then modeling of the jet
bending and Doppler boosting may allow one to reconstruct the 3D velocity and constrain the amount
of momentum flux injected into the jets (e.g., 
\citealt{2019MNRAS.484.4760B}
). 

The PWN inside the composite SNR CTA 1 
was observed with
CXO ACIS-S in 2003 for 49 ks (PI Halpern) and HRC-I in 2004 for 75 ks (PI Murray). The
high-resolution CXO image of the pulsar’s vicinity suggests a curved jet extending for $\approx$ 20$''$ south
of the pulsar and a hint of a faint compact torus \citep{2004ApJ...612..398H}. 
The radio and X-ray properties of SNR CTA 1 imply an age of 5-15 kyr
(\citealt{1993AJ....105.1060P}, \citealt{1997ApJ...485..221S}, \citealt{2004ApJ...601.1045S}), and
\cite{1993AJ....105.1060P} estimate a distance of $1.4\pm0.3$ kpc based on the neutral hydrogen observations.
The 316-ms radio-quiet PSR J0007+7303 (hereafter J0007) was discovered within the X-ray PWN with the
Fermi Gamma Ray Observatory's LAT instrument through its $\gamma$-ray pulsations \citep{2013ApJS..208...17A} followed by the detection of X-ray pulsations with
XMM-Newton \citep{2010ApJ...725L...6C}. The pulsar has a spin-down age of 14 kyrs, power $\dot{\rm E}$ = 4.5 $\times$ 10$^{35}$ erg/s, and
a relatively high magnetic field of 1.1$\times$10$^{13}$ G. 
The compact PWN spectrum was found to be rather hard, albeit uncertain, with photon index $\Gamma$ = 1–1.3 for the compact PWN and the jet \citep{2004ApJ...612..398H}, compared to those of the
majority of PWNe \citep{2008AIPC..983..171K}. 
\citealt{2013ApJ...764...38A} reported the discovery of TeV $\gamma$-ray emission from VER J0006+729,
coincident with the X-ray PWN. The energetics
and the relatively small
extent of the TeV emission (compared to the SNR)
support the PWN origin of the TeV photons as opposed to the SNR origin.
Most recently,
\citealt{2025SCPMA..6879503L}
detected significant ($17\sigma$) ultra-high energy (UHE) emission ($>$ 100 TeV) from 1LHAASO J0007+7303u
coincident with
J0007
and its PWN with a UHE source  extent smaller than that of the SNR.

\section{Observations and Data Reduction}
\subsection{Chandra CXO}

We 
use one archival (epoch 1)
and six new
Chandra observations (epoch 2)
of J0007
(see Table \ref{tab:cxo_obs} for details).
The data from these observations were reprocessed using the
{\tt chandra\_repro} tool
from the
Chandra Interactive Analysis of Observations (CIAO) software
package \citep{2006SPIE.6270E..1VF} version 4.15 with the Chandra
Calibration Database version 4.10.4. 

\begin{deluxetable*}{cccccc}
\tablecolumns{5}
\tablecaption{Summary of Chandra Observations of PSR J0007+7303
\label{tab:cxo_obs}
}
\tablewidth{0pt}
\tablehead{
\colhead{ObsID} & \colhead{Observation Date} & \colhead{Exposure (ks)} & \colhead{Instrument} & \colhead{Mode} 
}
\startdata
3835   & 2003 Apr 13   & 49.48 & ACIS-S & Very Faint \\
26662  & 2023 Aug 15   & 24.76 & ACIS-I & Very Faint \\
27102  & 2023 Dec 01   & 29.70 & ACIS-I & Very Faint \\
27103  & 2023 Nov 16   & 27.43 & ACIS-I & Very Faint \\
27104  & 2023 Dec 11   & 14.89 & ACIS-I & Very Faint \\
27105  & 2024 Aug 27   & 28.21 & ACIS-I & Very Faint \\
29118  & 2023 Dec 14   & 19.83 & ACIS-I & Very Faint \\
\enddata
\tablenotetext{}{}
{The data were collected using
Advanced CCD Imaging Spectrometer (ACIS; \citealt{2003SPIE.4851...28G}). For newer data 
I-array operated in the “Very Faint” timed exposure
mode was used while for the old data the back-illuminated ACIS-S3 chip with higher background was used in “Very Faint” mode. The observation
from 2003 is hereafter called epoch 1
and the observations from 2023/2024 are hereafter called epoch 2.
The total scientific exposure is 194.30 ks.}
\end{deluxetable*}

\subsection{Fermi-LAT}
We
use nearly 16 years of data
from the Fermi Gamma-ray 
Observatory's Large Area Telescope (LAT).
We select all events within a $10$\textdegree 
ROI centered around J0007 (RA=1\fdg7525, Dec=73\fdg0519).
Standard data quality cuts (DATA\_QUAL$>0$, LAT\_CONFIG==1) are applied with Fermitools {\tt gtmaketime}, and time intervals affected by bright solar flares or known anomalies are excluded. A zenith-angle cut of $<90$\textdegree suppresses Earth-limb contamination. Corresponding spacecraft files are used to generate live-time cubes and exposure maps (gtltcube, gtexpcube2). 
We use the selection of Pass 8 SOURCE class events, and the  galactic and isotropic diffuse models of \texttt{gll\_iem\_v07.fits} and \texttt{iso\_P8R3\_SOURCE\_V3\_v1.txt}.

For 
our analysis, we consider only the energy range of 50 GeV-1 TeV to 
avoid 
contamination of PWN spectrum by the pulsar, which has a cutoff of a few GeV \citep{2023ApJ...958..191S}. Although LAT sensitivity declines toward 1 TeV, the point-spread function improves dramatically above $\sim$10 GeV ($\leq0.1$\textdegree), allowing for morphological studies of extended emission and extension fitting. Beyond the Galactic and isotropic diffuse templates, we 
include all sources from the Fermi LAT Fourth Source Catalog Data Release 4 \citep{2023arXiv230712546B} within 15 degrees of J0007. The central pulsar, J0007, is removed from the model - replacing it with a test source modeled by a power-law 
because the pulsar contribution above 50 GeV is negligible and we expect to see only the PWN emission.
The spectral modeling consists of optimizing the ROI by fitting the sources in an iterative manner with {\tt gta.optimize()}.   For all sources but the PWN we only free normalizations and keep the other spectral parameters at their best-fit values from the 4FGL catalog. 

Spectral and spatial modeling of the 
PWN is done via binned likelihood analysis ({\tt gtlike} task within the Fermipy package; \citealt{2017ICRC...35..824W}). Counts maps and model maps are constructed with fine spatial binning ($0.05$\textdegree per pixel).  Extension is tested by comparing a point-source hypothesis against extended morphologies (e.g., 2D Gaussian or disk templates), with significance expressed as TS$_{\rm ext} = -2\Delta \log \mathcal{L}$ where $\mathcal{L}$ is the 
likelihood function.

\section{Results}

\subsection{Pulsar Motion\label{sec:pulsar_motion}}

All observations from epoch 2 are collected within a one year period, and the time interval between epoch 1 and epoch 2 is about 20 years.
To measure the pulsar motion we performed a relative astrometric correction
by matching X-ray sources seen
in the images from epoch 2 to those in the image from
epoch 1.

The proper motion of the pulsar is then computed between 
epoch 1 and the
merged epoch 2 image produced from individual images after co-aligning them with the epoch 1 image.
 We used the latter as the  reference because it is fairly long and the sensitivity of ACIS
 was significantly 
 greater at that time\footnote{\href{https://cxc.harvard.edu/cal/Acis/Cal_prods/qeDeg/index.html}{https://cxc.harvard.edu/cal/Acis/Cal\_prods/qeDeg/index.html}. }. The details about the co-alignment procedure can be found in Appendix \ref{sec:astrometry}. 

To determine pulsar position and its uncertainty 
we 
binned the
images by a factor of 0.2 and
defined a square grid with the spacing of $0.025''$ 
and overall size of  2$''$ $\times$ 2$''$  centered on the brightest pixel in each image
of the pulsar.
For the centroiding procedure, circular regions of different sizes 
(radii ranging from $0.25''$ to $0.5''$)
are then placed at each grid point and the sum of photon counts within each circle is calculated using CIAO's task {\tt dmstat}.
The
grid point corresponding to the
region with the maximum sum
for each 
circle radius
is taken as the pulsar position.
The
pulsar positions from the two different epochs 
were then used to calculate the proper motion (PM).

The center of the reference (epoch 1) observation was found to be 
$(\mathrm{RA}, \mathrm{Decl.}) = (00^{\mathrm h}\,07^{\mathrm m}\,01.582^{\mathrm s} \pm 0.010^{\mathrm s},
+73^\circ\,03'\,08.238'' \pm 0.055'')$,
and for the merged, epoch 2 image it was 
$(\mathrm{RA}, \mathrm{Decl.}) = (00^{\mathrm h}\,07^{\mathrm m}\,01.645^{\mathrm s} \pm 0.017^{\mathrm s}, 
+73^\circ\,03'\,08.371'' \pm 0.069'')$.
The uncertainty, $\sigma_{\rm stat}$, was taken as the standard deviation 
of the positions obtained by summing counts with different circle radii (see above).
The PM was found to be 12 $\pm$ 3 $\pm$ 11 mas yr$^{-1}$ in R.A. and 6 $\pm$ 5 $\pm$ 14 mas yr$^{-1}$ in Decl. where the first error is $\sigma_{\rm stat}$
and the second error is $\sigma_{\rm sys}$
(see Appendix \ref{sec:astrometry} for description of $\sigma_{\rm sys}$).
This corresponds to an upper limit on the transverse velocity of 100 $\pm$ 20 $\pm$ 110 km s$^{-1}$, at the distance of 1.4 kpc. 

\subsection{Imaging Analysis}

\begin{figure*}
    \centering    \includegraphics[width=1.0\linewidth]{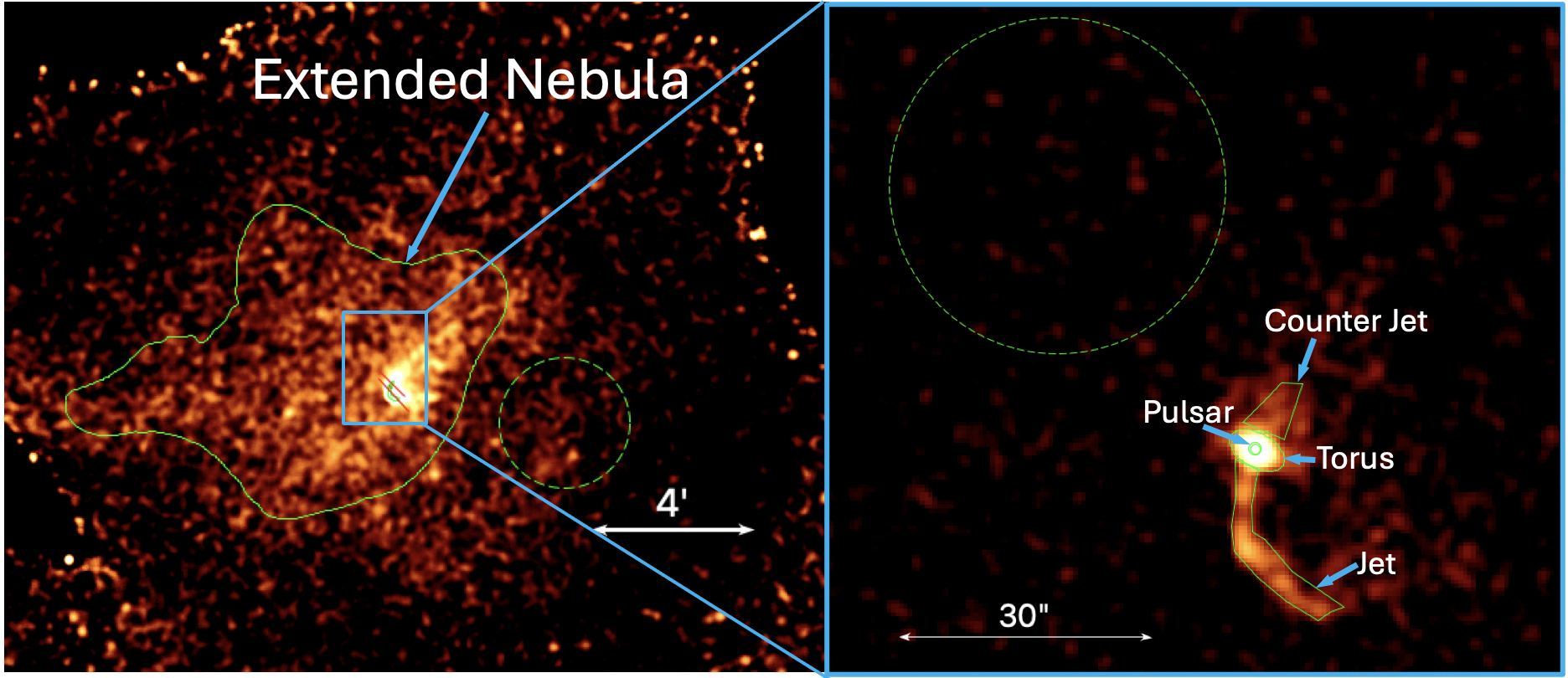}
    \caption{Regions used for spectral extraction. Dashed green lines indicate background regions. The left panel shows the merged, exposure corrected epoch 2 image with point-sources subtracted, binned by a factor of 2 and smoothed by a Gaussian. The right panel is unbinned and smoothed by a Gaussian.}
    \label{fig:cxo_reg}
\end{figure*}

Figures \ref{fig:cxo_reg} 
and \ref{fig:surf_bri}
show merged
images
of J0007 and its 
PWN.
On smaller scales (right panel of Fig. \ref{fig:cxo_reg}), the most prominent PWN feature is the $\sim20''$-long
jet extending south of the pulsar 
and sharply bending to the west.
There is also a hint of 
much fainter, and possibly wider,   counter-jet
to the north of the pulsar.
The central region around the pulsar appears elongated, roughly perpendicular to
initial segments of the jet/counter-jet, 
suggesting the presence of a compact torus.
On larger, arcminute scales 
the diffuse emission image
(left panel of Fig. \ref{fig:cxo_reg})
shows
clear evidence of
PWN emission 
seen primarily east and north up to $\sim8'$ from the pulsar.

\begin{figure}
    \centering
\includegraphics[width=1.0\linewidth]{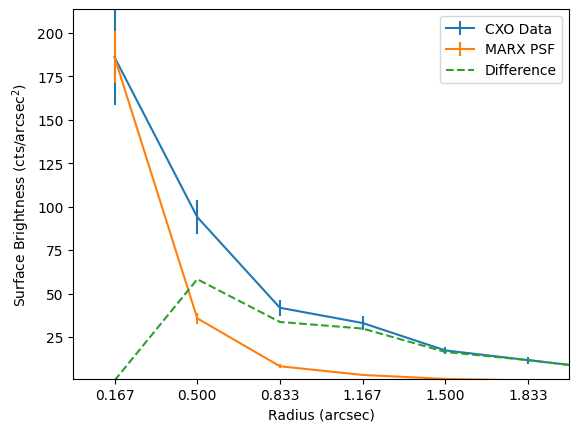}
    \caption{Radial profile of the CXO image compared to the MARX simulation. The green dashed line shows the difference between the data and the simulated PSF.}
    \label{fig:rad_prof}
\end{figure}

\begin{figure}
    \centering
\includegraphics[width=1.0\linewidth]{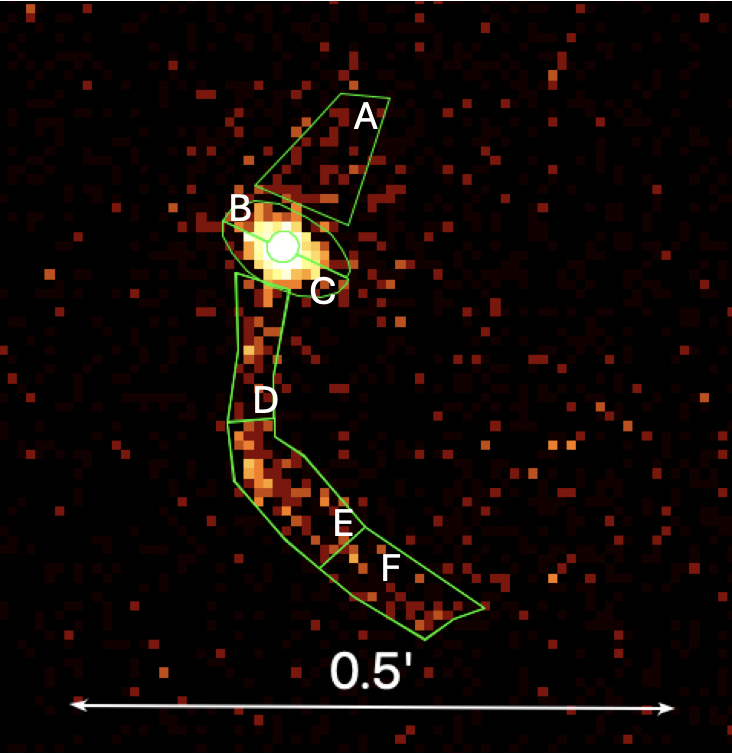}
    \caption{Unbinned (pixel size 0.5'') image of all CXO observations aligned and merged. In green are the regions used for examining surface brightness
    }
    \label{fig:surf_bri}
\end{figure}

To quantify the presence  of the extended emission in the vicinity of the pulsar, we constructed a radial profile. 
We perform a 
point-source simulation 
using MARX
\footnote{https://space.mit.edu/cxc/marx/}
\citep{2012SPIE.8443E..1AD}
to model the point-spread function (PSF)
on the detector at the pulsar's location.
We compare the extension of the simulated point source to the extension of the emission surrounding the pulsar in the 
merged image of the epoch 2 CXO observations 
after alignment (see Appendix \ref{sec:astrometry} for description of alignment) in Figure \ref{fig:rad_prof}. 
We use circular annuli with a width of $0.33''$ 
centered on the brightest pixel (with the size of $0.5''$) in both the simulated image and the merged 
CXO image.
The simulation profile is 
normalized to match the value in the brightest pixel in the 
actual observation image.

In the same figure we also plot the difference between the simulated and observed profiles. 
One can see
a clear excess in surface brightness in the CXO data 
with a maximum deviation near $r\approx0.5''$. We attribute this excess to the presence of the torus.
Comparing the profiles, 
we also find that only $\approx$12\% of the photons
between $\sim 0.8''-1.5''$ 
can be attributed to the pulsar. 

We also examined the 2D distribution of the background-subtracted surface brightness 
for different parts
of the PWN (see Fig. \ref{fig:surf_bri}). 
The jet (regions D, E, and F) is brighter than the counter jet (region A) by a factor of 1.42, 
with surface brightnesses of 4.51$\pm$0.29 cts arcsec$^{-2}$ and 3.18$\pm$0.40 cts arcsec$^{-2}$, respectively. 
From Figure \ref{fig:surf_bri} one can see that the jet surface brightness varies significantly along its extent with the jet being fainter near the
pulsar and brightest at the place where it starts to bend significantly.
The background-subtracted surface brightnesses for the jet's regions D, E, and F are $(4.84\pm0.59)$, $(5.59\pm0.50)$, and $(3.36\pm0.44)$ cts arcsec$^{-2}$, respectively.

Similar to the jet and counter-jet, the southern-eastern (SE) side of the torus (region C) is  brighter than the north-western (NW)  side (region B) by a factor of 1.3,
with background subtracted surface brightnesses of (24.31$\pm$1.55) cts arcsec$^{-2}$ and (18.51$\pm$1.38) cts arcsec$^{-2}$, respectively.
Assuming that the torus appears to be more extended (NE-SW) in one direction than in the other (NW-SE)  due to the tilt of its axis (which is also the pulsar spin axis) with respect to the line-of-sight, the tilt (viewing angle $\zeta$)  can be inferred by fitting an ellipse to the torus and measuring the ratio of the ellipse axes. To accomplish this, we fit an half-ellipse  to the brighter, SE part of the torus. We first remove the pulsar by applying a gradient filter (Sobel filter from {\tt scikit-image} python package) to an image of compact nebula smoothed with $r=0.3''$ Gaussian. This effectively  enhances the outer edge of the torus while suppressing the pulsar itself. We then construct the ridge curve (tracing the maximum of brightness in the gradient image) by simply finding the maxima for each column  in the rotated image with longest torus dimension being along the rows. These points are then mirrored to create a complete torus, which is  fitted using the {\em EllipseModel()} function from the {\tt scikit-image}  library 
\citep{vanderWalt2014}. The cosine of the 
viewing angle is 
the ratio of the semi-minor and semi-major axes of the best-fit ellipse, which gives 
$\zeta\approx50^{\circ}$.

\subsection{Spectral Analysis}
\label{sec:spectral}

We performed a spectral analysis of J0007 and its 
compact
PWN
simultaneously fitting spectra from all CXO observations
(see Table \ref{tab:cxo_obs}).
The spectral fits were performed in Sherpa (v.4.15.0;
\citealt{2001SPIE.4477...76F}) using the XSPEC's
power-law (PL) model modified by interstellar extinction according to  {\tt tbabs} model
(the absorption cross
sections and abundances are described in 
\citealt{2000ApJ...542..914W}).
The energy range for the fits was restricted to 0.5-7 keV, and the spectra were then jointly fit
using all observations.
The hydrogen column density, $N_{\mathrm{H}}$,
is fixed at the value obtained from the jet's  spectrum fit (see below) because it has relatively low background contribution and is not affected by a thermal component (which may be present in pulsar's spectrum).
The best-fit model parameter uncertainties are given at the 68\%
(1$\sigma$) confidence level.
Figure \ref{fig:cxo_reg} shows all regions used for spectral extraction. 
For each observation (after co-alignment) spectra are extracted from the same regions on the sky  
for the pulsar, torus, jet, counter-jet, and extended nebula regions. 
The pulsar spectrum is extracted from a circular region 
(R.A.\ = 0$^{\rm h}$ 07$^{\rm m}$ 01$^{\rm s}$.6028
, Dec.\ = 
+73$^\circ$ 03' 08$''$.285, r= 0.6$''$),
which is acceptable for both epochs given the small magnitude of the pulsar proper motion (see Section \ref{sec:pulsar_motion}).
Based on the radial profile, $\sim80\%$ of counts in this region can be attributed to the pulsar.
The torus region is defined as an ellipse
(centered at R.A. =
0$^{\rm h}$ 07$^{\rm m}$ 01$^{\rm s}$.5688
, Dec. = 
+73$^\circ$ $03'$ $08.133''$, with r$_1$ =$3.56''$ , r$_2$= 2.04$''$
oriented at
$\theta$ = 240.55$^{\circ}$ East of North). To minimize contamination from the pulsar, an $r=0.8''$ circular region
centered on the pulsar
was excluded. 
We estimate that $\approx 10\%$ of the counts in this region come from the pulsar.
The extended nebula region was created by following the constant brightness contours created  in SAO DS9\footnote{https://sites.google.com/cfa.harvard.edu/saoimageds9}.

 For fits of the pulsar, torus, and jet/counter-jet regions 
 we take the background from
 a circular region
 with r = $20''$ 
 and for the extended nebula we use
 an r = $100''$ radius,
 shown by dashed lines in
 the respective panels of
 Figure \ref{fig:cxo_reg}. We use the $\chi^2$ statistic for fits to all spectra except for the counter-jet spectrum where we use {\em wstat\footnote{CIAO implementation of cash statistic with background included.}}. 
 
\begin{figure}[h!]
    \centering
\includegraphics[width=0.76\linewidth]{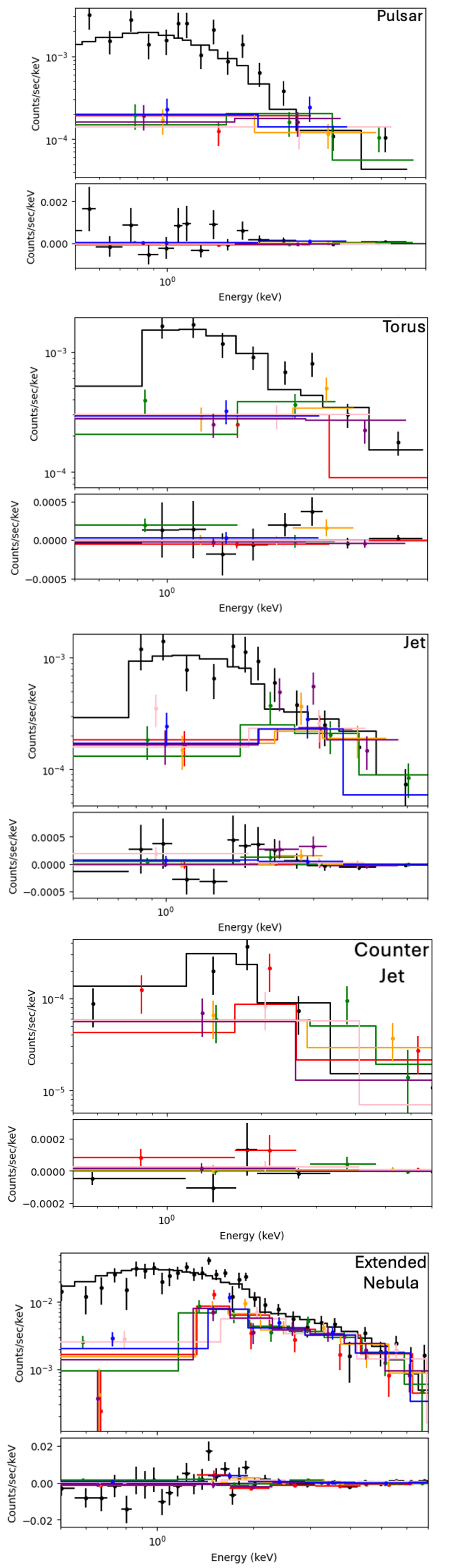}
    \caption{CXO spectral fits.}
    \label{fig:cxo_spectra}
\end{figure}

To fit the pulsar spectrum (shown in Fig. \ref{fig:cxo_spectra}), we first grouped
counts by requiring $\geq$9 counts per energy bin. The absorbed PL model provides a good fit (see Table  \ref{tab:fit_stats} for details)
with best-fit PL $\Gamma$ =
2.42 $\pm$ 0.18. The corresponding absorbed 
flux is (1.83 $\pm$ 0.26) $\times$
10$^{-14}$ erg cm$^{-2}$ s$^{-1}$.
The unabsorbed luminosity  is
6.47 $\times$ 10$^{30}$ erg s$^{-1}$ at  d = 1.4 kpc.

The torus spectrum 
(grouped by $\ge20$ counts per bin) 
is much harder
with $\Gamma = 1.34 \pm 0.11$ and an
absorbed flux of $(3.00\pm 0.40) \times 10^{-14}~\mathrm{erg~cm^{-2}~s^{-1}}$, yielding an unabsorbed luminosity of $8.10 \times 10^{30}~\mathrm{erg~s^{-1}}$.

The jet region's (regions D+E+F in Fig. \ref{fig:surf_bri}) spectrum (grouped by $\ge10$ counts per bin) is 
also quite hard, characterized by $\Gamma = 1.25 \pm 0.20$
and
an absorbed flux of $(2.28\pm0.37) \times 10^{-14}~\mathrm{erg~cm^{-2}~s^{-1}}$, corresponding to an unabsorbed luminosity of $6.10 \times 10^{30}~\mathrm{erg~s^{-1}}$.
The hydrogen column density was fit for this region 
and was found to be $N_{\mathrm{H}}=(3.1\pm1.7)\times10^{21}$ cm$^{-2}$.

Due to the faintness of the counter-jet, its emission is largely buried under the high ACIS-S3 background in ObsID 3835. 
Therefore, its spectrum 
was extracted from 
only the new 
data obtained with ACIS-I.
This region (fit with an absorbed PL model using the wstat
) displays a softer spectrum than the jet with $\Gamma = 1.49 \pm 0.25$
with an absorbed flux of $(5.51\pm1.3) \times 10^{-15}~\mathrm{erg~cm^{-2}~s^{-1}}$, 
with unabsorbed luminosity of $1.53 \times 10^{30}~\mathrm{erg~s^{-1}}$.

The combined unabsorbed luminosity of compact nebula comprised of all of the above-mentioned components (except for the pulsar) is 1.57$\times10^{31}\,\mathrm{erg~s^{-1}}$.

The fainter extended nebula emission
is also
likely lost 
in the high ACIS-S3 background in ObsID 3835,
so this observation is not included when fitting the spectrum from this region. 
The fit of the spectrum (grouped by $\ge400$ counts per bin) 
yields $\Gamma = 1.85 \pm 0.11$
and the absorbed flux is $(1.10\pm0.13) \times 10^{-12}~\mathrm{erg~cm^{-2}~s^{-1}}$, leading to an unabsorbed luminosity of $3.29 \times 10^{32}~\mathrm{erg~s^{-1}}$.

\begin{deluxetable*}{ccccccc}
\tablecolumns{7}
\tablecaption{Spectral Fit Results for PWN Regions and Point Sources
}
\tablewidth{0pt}
\tablehead{
\colhead{Region} & \colhead{Area} & \colhead{Counts (bkg \%)} & 
\colhead{$\Gamma$} & \colhead{$\mathcal{N}_{-6}$} & 
\colhead{Fit Statistic (d.o.f)} & \colhead{$F_{-14}^{\rm unabs}$}
}
\startdata
Pulsar & 0.97 & 252 (0.3\%) & $2.42 \pm 0.18$ & $8.05 \pm 0.83$ & 38.21 (25) & $2.76\pm0.33$ \\
Torus & 21.30 & 459 (3.6\%) & $1.34 \pm 0.11$ & $4.81 \pm 0.49$ & 40.52 (39) & $3.45\substack{+0.46 \\ -0.39}$ \\
Jet & 64.15 & 338 (14.6\%) & $1.25 \pm 0.20$ & $3.29 \pm 0.79$ & 23.41 (28) & $2.60\substack{+0.43 \\ -0.37}$ \\
Counter Jet$^\dagger$ & 22.27 & 58 (14.3\%) & $1.49 \pm 0.25$ & $1.05 \pm 0.25$ & 102.85 (83) & $0.65\substack{+0.14 \\ -0.12}$ \\
Extended Nebula & 163,094 & 52,030 (84.1\%) & $1.85 \pm 0.11$ & $300.26 \pm 29.22$ & 87.01 (74) & $140.38\substack{+16.25 \\ -14.46}$ \\
\enddata
\vspace{-0.5cm}
\tablenotetext{}{}
{Region area in units of arcs$^2$, net counts (background counts as a percentage of net counts; Counter Jet and Extended Nebula region information is taken only from observations in epoch 2), photon index $\Gamma$, normalization $\mathcal{N}_{-6}$ in units of $10^{-6}$ photons s$^{-1}$ cm$^{-2}$ keV$^{-1}$ at 1 keV, fit statistic and degrees of freedom (d.o.f), and unabsorbed flux $F_{-14}^{\rm unabs}$ in units of $10^{-14}$ erg s$^{-1}$ cm$^{-2}$ in the 0.5–7 keV band. Flux uncertainties are 1$\sigma$. Fits use chi-squared statistics except for the Counter Jet region which uses wstat. $N_{\mathrm{H}}$ is fit in the jet region and frozen to this value ($N_{\mathrm{H}}$=0.31$\times10^{22}$ cm$^{-2}$) when fitting spectra from other regions.}
\label{tab:fit_stats}
\end{deluxetable*}

\subsection{Fermi Spatial and Spectral Analysis}
\label{sec:fermi}
Prior to the spectral fitting of the J0007 source, we test several spatial models (a point source, radial disk, and radial gaussian) to evaluate the model that is most likely representative of J0007 using a likelihood analysis \citep{1983ApJ...272..317L}. The test statistic (TS) is defined as the logarithmic difference between the likelihood of the source case and the likelihood of the null hypothesis (no source): 

$$TS=2[\log\mathcal{L}({\rm source})-\log\mathcal{L}({\rm null})]\quad,$$
where $\mathcal{L}$ is the Poisson likelihood of the data given the model. Here, the TS is approximately distributed as $\chi^2$ such that such that  
$\sqrt{\rm TS} = \sigma$ provides an estimate of the detection significance. We detect J0007 with a significance of TS = 53.71 corresponding to a significant $\gamma-$ray excess best modeled with a Radial Disk (RD) centered around the pulsar. 

\begin{figure}
    \centering
    \includegraphics[width=\linewidth]{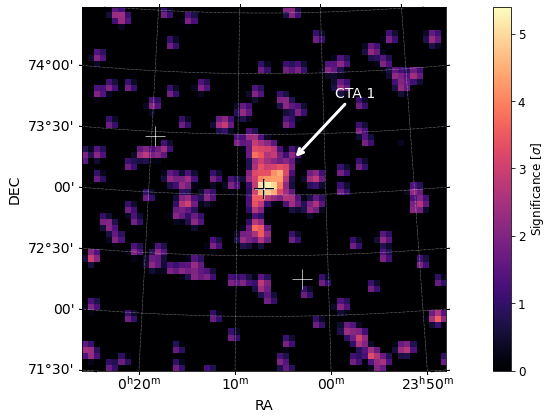}
    \caption{
    TS map of the J0007 field derived from Fermi-LAT data. The color scale indicates detection significance defined as $\sqrt{TS}$.
    The extended emission associated with the J0007 PWN is clearly visible near the center of the field, highlighted by the white arrow.}
    \label{fig:TSmap}
\end{figure}

\begin{figure}
    \centering
    \includegraphics[width=\linewidth]{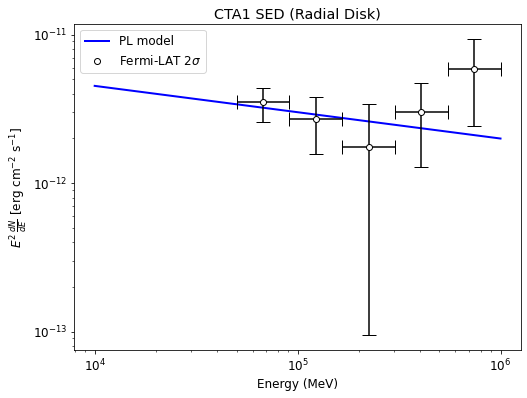}
    \caption{Spectral energy distribution (SED) of the J0007 PWN assuming a radial 
    disk (RD) spatial model. Black points show the Fermi-LAT flux measurements with a TS$>4$, 
    corresponding to a $2\sigma$ measurement. The solid blue line represents the best-fit 
    PL model.}
    \label{fig:fermi_sed}
\end{figure}

We do not attempt to localize the source and find that the RD has evidence for extension TS$_{\rm ext} = 20.21$ with a radius of $r=0.32$\textdegree$\pm0.08$. A TS map of the J0007 PWN is shown in Figure~\ref{fig:TSmap} with the Radial Disk best-fit model removed. When the model is introduced, the TS map surrounding J0007 becomes consistent with the background. From this best-fit spatial model, the spectrum from 50 GeV–1 TeV is best characterized by a simple power-law model, consistent with inverse Compton (IC) emission from the highest-energy electrons. The best-fit photon index is $\Gamma_\gamma=2.18\pm0.29$. The spectral energy distribution (SED) is extracted in five logarithmically spaced energy bins across this band, and is shown in Figure~\ref{fig:fermi_sed}.

\section{Discussion}
The  
combined CXO ACIS images (shown in Figure \ref{fig:cxo_reg}) of J0007 and its PWN reveal 
a clear jet component extending $\approx20''$ to the SE of the pulsar and bending to the SW, 
and a faint counter jet extending to the NW.
The emission near the pulsar appears to be  elongated 
in the NE/SW directions, roughly perpendicular to the jet/counter jet
suggesting the presence of a torus.
This is confirmed by the clear excess in surface brightness over the point-spread function
past $\sim0.8''$ shown 
in Fig. \ref{fig:rad_prof}.
Figure \ref{fig:cxo_reg} also 
shows that the compact PWN is embedded within a
large-scale diffuse emission. 

\begin{figure}
    \centering    \includegraphics[width=1.0\linewidth]{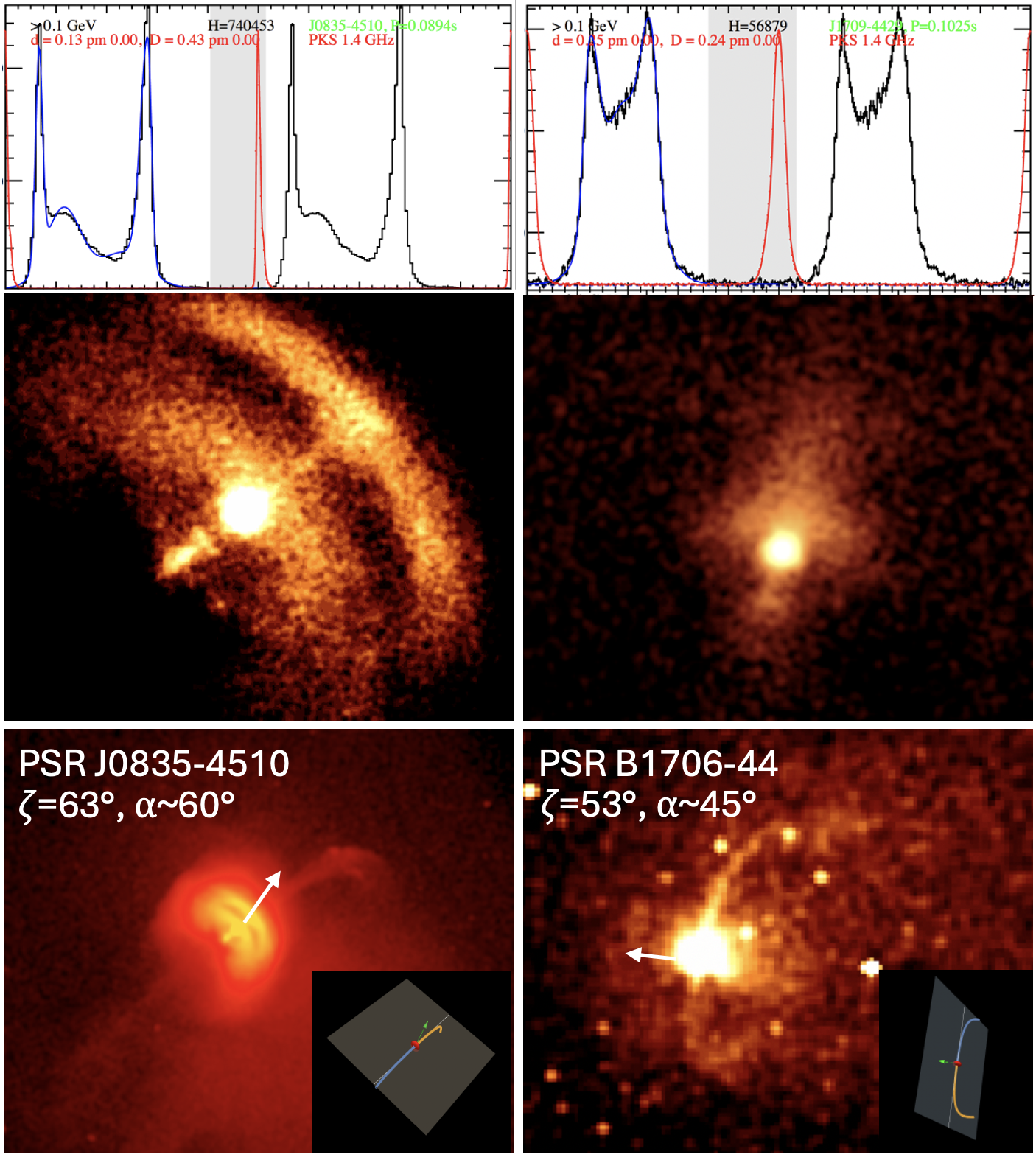}
    \caption{
    $\gamma$-ray and radio 
pulse profiles 
(top row, from \citealt{2013ApJS..208...17A})
of pulsars with relatively small and compact tori and large jets. The images are from {\sl CXO} ACIS observations. 
See, e.g.
\citealt{Liu2024VelaPWN},
\citealt{2021ApJ...908...50D}
for each source, respectively. 
The second row shows the small-scale jet/torus structures
and the third row is a zoomed out image to show the effects of ram pressure on the large scale structure.
The white arrow shows the previously suggested direction of proper motion
(see 
\citealt{2003ApJ...596.1137D},
\citealt{2021ApJ...908...50D}).
The insets illustrate possible orientations for each of the pulsars' spin axes with respect to the velocity vector and the line of sight
computed using equation 22 in \citealt{2011MNRAS.412.1870P}.
The curves represent two jets bent by the ram pressure due to the pulsar motion.
}
    \label{fig:pulse_profs}
\end{figure}

The bending of the jet could be caused by ram pressure if J0007 has a significant velocity with respect to the surrounding medium.
The age estimates and the offset of 
the pulsar from the geometrical center of the radio
SNR suggests pulsar's transverse velocity, v$_{\perp}$ = 450–1,500 km/s for $d = 1.4$ kpc. The range is
large due to the uncertain estimates of the pulsar's age and birth place within the SNR.
This velocity is barely compatible with the $3\sigma$ upper limit on the transverse velocity of the pulsar (see Section \ref{sec:pulsar_motion}) suggesting that 
the SNR is older or the center of the SNR is even closer to the pulsar than assumed. The latter could happen due to SNR expanding more rapidly on one side than on the other. 
It is also possible that the bending of the jet is caused by interaction with the SN reverse shock,
which would increase the velocity of the oncoming medium in the frame of the pulsar
without 
requiring higher proper motion measured with respect to background sources.

The jet shows significant variation in brightness along its extent, brightening notably as it bends to the SW (see region E in Fig. \ref{fig:surf_bri}).
This 
is consistent  with ram pressure pushing
the jet
backwards.
This can
compress the jet,
leading to an increased magnetic field,
and thus an increased surface brightness.
A similar brightening of the forward (pointing in the direction of pulsar's  motion)  jet  is seen in 
the Vela pulsar 
and PSR B1706--44
(hereafter B1706, 
see 
Fig. \ref{fig:pulse_profs}). For both of these pulsars proper motion has been accurately measured
(\citealt{2003ApJ...596.1137D}, \citealt{2021ApJ...908...50D}).  
For the Vela pulsar, on small scales (second row of Fig.\ref{fig:pulse_profs}) 
Doppler 
boosting can be invoked to explain the
brightening in the
NW part of the torus and the SE inner (small-scale) jet 
if the projection of the flow velocity component onto the line-of-sight is directed toward the observer.
On larger scales (third row of Fig. \ref{fig:pulse_profs}), however, the NW jet is clearly brighter despite
the expected de-boosting.
This could be explained by the interaction with the 
oncoming medium 
which  bends the jet west of its original direction. 
A similar brightening behavior can be seen in the jet/counter-jet of B1706.

In the J0007 PWN the torus also shows significant variation in surface brightness, 
with the SE side being brighter than the NW side (see regions C and B in Fig. \ref{fig:surf_bri}, respectively).
This implies that the pulsar's
equatorial
axis is tilted 
relative to the 
line-of-sight 
with the SE side 
being closer to the observer
and thus experiencing Doppler boosting.
We use the ellipticity of the torus
to measure the
viewing angle, $\zeta$, between the pulsar spin axis and the line of sight,
which is found to be 
$\zeta\approx50^\circ$.
This allows us
to 
constrain
the magnetic inclination angle, 
$\alpha$,
using 
pulse-profile modeling
results from \citealt{2009ApJ...695.1289W}.
The viewing angle that we measure 
is roughly compatible with the predictions,
for the pulsar's GeV lightcurve,
of both the Outer Gap
and Two-Pole Caustic (Slot Gap) 
models of magnetospheric emission
for $\alpha\approx20^\circ$-- $70^\circ$.
The above inferred value of $\alpha$ is also compatible with the 
modeling by 
(\citealt{2024A&A...687A.169P},
Figure A.2) 
using a force-free striped-wind model
\citep{2002A&A...388L..29K}
which suggests
that $\alpha\sim45^{\circ}$ for $\zeta\sim50^{\circ}$.

\begin{figure}
    \centering
\includegraphics[width=1.0\linewidth]{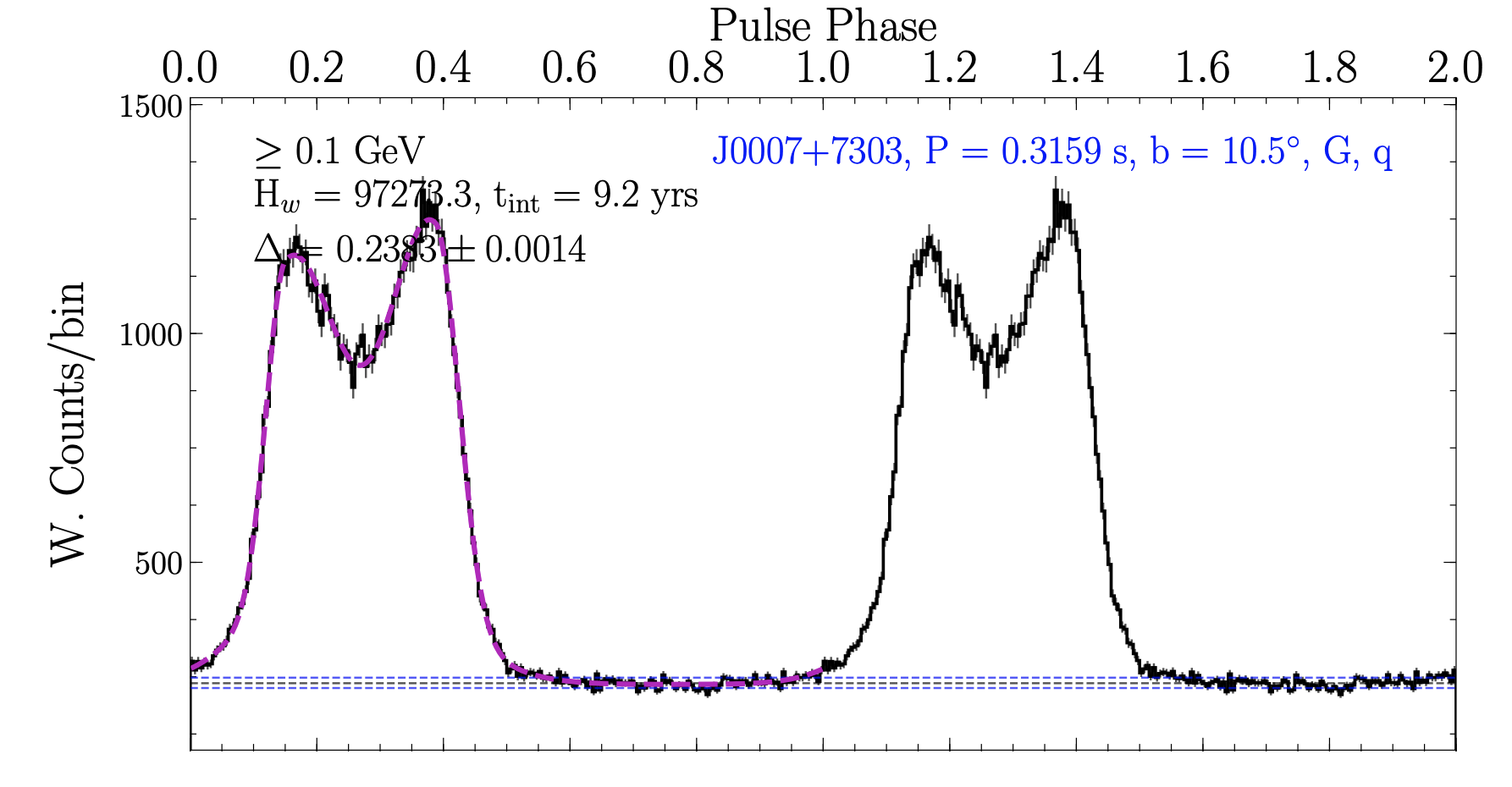}
    \caption{Fermi GeV pulse profile of J0007 \citep{2023ApJ...958..191S}}
    \label{fig:CTA1_pulse_prof}
\end{figure}

We then compare 
the surface brightness 
of the inner jet, measured before noticeable compression effects 
to
that of
the torus
(regions D and B+C from Fig. \ref{fig:surf_bri}, respectively)
and find that
the torus 
is brighter.
Relativistic magnetohydrodynamics simulations 
(see e.g. \citealt{2016MNRAS.462.2762B}, \citealt{2017SSRv..207..137P})
suggest that as the magnetic inclination angle
increases the average surface brightness of the torus relative to the jet will also increase making it more luminous.
Since the initial  wind magnetization, $\sigma$, also affects the  surface brightness according to  simulations (see Figures 7 and 8 in \citealt{2016MNRAS.462.2762B}),
this implies
for torus-dominated PWNe that 
the magnetic inclination angle is not too small
($\alpha \gtrsim 20^\circ$). This 
condition is compatible with the
requirements for $\zeta$ and $\alpha$
from magnetospheric emission models to reproduce the observed $\gamma$-ray pulse profile (see Figure \ref{fig:CTA1_pulse_prof}).

It is worth noting that the pulsars shown in Figure \ref{fig:pulse_profs}  share similar pulse profile and compact PWN structures  with those of J0007. This may not be  a coincidence if these structures are imprinted by the geometry defined by $\alpha$ and $\zeta$. Indeed, equation 22 from  \citep{2011MNRAS.412.1870P} gives $\alpha$ and $\zeta$ values close to those inferred for J0007 and its PWN (see Fig.~\ref{fig:pulse_profs} ).

All of the components of the J0007 PWN (Torus, Jet, and Counter Jet) display surprisingly hard spectra (see Table \ref{tab:fit_stats}),
indicating 
that there is minimal radiative cooling in the vicinity of the pulsar
and we are seeing the spectrum of the particles produced by the acceleration mechanism at work.
These spectra can be used to infer the dominant acceleration mechanism by comparing to results from theoretical modeling.
Within the reconnecting striped wind models 
(e.g. \citealt{1999A&A...349.1017B})
the larger $\alpha$ is the larger the volume 
occupied by the striped pulsar wind zone, which is
where magnetic field energy can be converted into the kinetic energy of
pulsar wind particles.
\cite{2014ApJ...783L..21S} show that increased spectral hardness
(photon index $\lesssim$1.5) implies that magnetic reconnection is the dominant particle acceleration mechanism.
The combined luminosity 
(in the 0.5-7 keV energy band)
of the compact PWN (Torus + Jet + Counter Jet) is 1.57$\times10^{31}$ erg s$^{-1}$, 
resulting in a radiative efficiency of $3.5\times10^{-5}$,
which is on the lower end for PWNe (see e.g. \citealt{2008AIPC..983..171K},
\citealt{2008ApJ...682.1166L},
\citealt{2010AIPC.1248...25K}).
Therefore, it is possible that 
the 
hard PWN spectrum,
the  UHE detection by LHAASO, and 
the low PWN X-ray radiative efficiency
are indicating that 
the particles that make it to the striped wind zone are accelerated to very high energies,
but the number of accelerated particles is relatively low.
This would be the case if this pulsar had a low pair multiplicity.
\cite{2019ApJ...871...12T} show that the pair multiplicity is maximized for pulsar with 
a magnetic field $4\times10^{12}\lesssim B \lesssim 10^{13}$ G and surface temperatures $\gtrsim10^6$ K.
Since J0007 has a magnetic field slightly above this range and a surface temperature of 
$<6.6\times10^5$ K 
\citep{2004ApJ...612..398H} it is possible that the pair multiplicity is indeed low.

\subsection{SED Modeling}

\begin{figure*}
    \centering
    \includegraphics[width=1\linewidth]{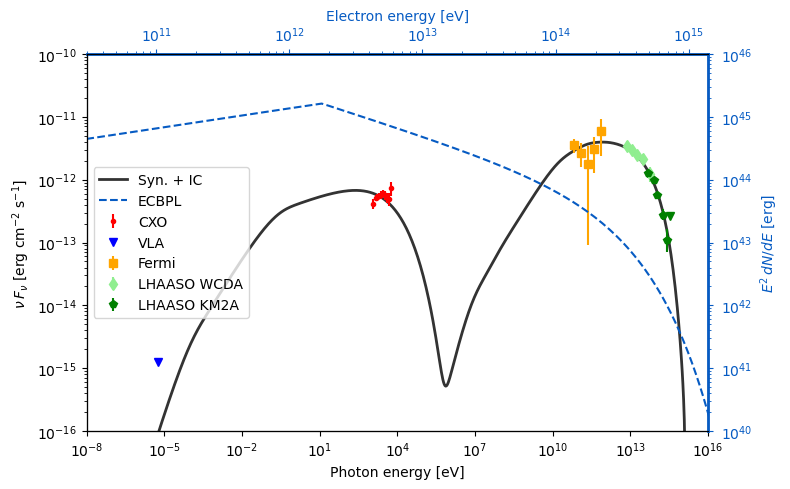}
    \caption{Broadband SED of the J0007 PWN modeled with Naima. The data shown include the VLA upper limits (ULs) \citep{2013ICRC...33.2656G}, CXO data (this work), Fermi (this work),
    and LHAASO \citep{2025SCPMA..6879503L}. The sum of the synchrotron and IC radiative models is shown by the solid black line, corresponding to the bottom and left axes. The input ECBPL electron population model for the radiative models is shown by the blue dashed line, corresponding to the right and top axes.}
    \label{fig:sed}
\end{figure*}

Since this is a young pulsar with a comparable PWN extent in X-rays and $\gamma$-rays 
($\sim$0.1$^\circ$ and $\sim$0.3$^\circ$, respectively)
we first attempt to model the multi-wavelength spectrum with a simple one-zone leptonic radiation model 
(see Fig. \ref{fig:sed}). 
We use the python package {\tt naima} \citep{2015ICRC...34..922Z} to model the electron SED with an exponential cutoff broken power-law (ECBPL)

\begin{align}
    f(E) =& \exp\left(-\left(\frac{E}{E_{\text{c}}}\right)^{\beta}\right) \times \notag \\
&\begin{cases}
A \left(\frac{E}{E_0}\right)^{-p_1}, & E < E_{\text{b}} \\
A \left(\frac{E_{\text{b}}}{E_0}\right)^{p_2 - p_1}
\left(\frac{E}{E_0}\right)^{-p_2}, & E > E_{\text{b}}
\end{cases}
\end{align}
where 
$E_{\rm c}$ is the cutoff particle energy,
$\beta$ is
the cutoff exponent,
$A$ is the normalization (in units of particles eV$^{-1}$ at
$E_b$),
$E_b$ is the break energy,
$E_0=1$ erg is the reference energy, 
$p_1$ is the slope of the
electron SED before the break,
and $p_2$ is the slope of the
electron SED after the break.
We chose a broken PL because for an extended PWN region 
we expect a cooling break to develop in the spectrum. We note that this model is not the injection  particle SED but the evolved one.
Using this particle SED we calculate the radiation spectrum produced by synchrotron and IC processes,
which depend on the magnetic field and 
the energy density of the seed photon fields,
in this case fields are added for the CMB and ambient starlight (T$=4000$ K, $u_E=0.5$ eV/cm$^3$, where T is the temperature and $u_E$ is the energy density) since the PWN is near the Galactic plane. 
This model is then fit to the data using an MCMC sampler, where $E_{\rm c}$, $A$, $E_{\rm b}$, $p_1$, and magnetic field are the free variables.
The slope of the electron SED after the break is constrained by $p_2=p_1+1$ as expected from synchrotron cooling (with $p=2\Gamma-1$).
This behavior is also confirmed observationally by the difference between the
extended nebula (cooled) and
torus (uncooled) 
spectra ($\Delta\Gamma=1.85-1.34\approx0.5$; see Table \ref{tab:fit_stats}). 
The minimum and maximum particle energies used in the radiative models are 30 GeV and 1.4 PeV, respectively, 
where the minimum energy is chosen 
to fit the VLA upper limit
and the maximum energy corresponds to $E_{\rm PC}$, the total potential drop energy across the NS polar cap.

Previously published VERITAS data \citep{2013ApJ...764...38A}
was not included in the fit, as the points are discrepant with the LHAASO WCDA data,
likely caused by an underestimated integration area used in the VERITAS analysis.
The best-fit cutoff energy was found to be 
$E_{\rm c}=
(0.21^{+0.39}_{-0.01})$ PeV.
This
cutoff energy 
corresponds to $\sim$15\% of the current $E_{\rm PC}$
which is unusually high according to some simulations \citep{2023ApJ...943..105H}.
It may be that $E_{\rm PC}$ was significantly larger when the NS was accelerating particles to 0.21 PeV,
as it is proportional to $\sqrt{\dot{E}}$, which can change 
significantly throughout the lifetime of a  pulsar.
This would lead to the cutoff energy representing a smaller fraction of $E_{\rm PC}$ in the past.
This may 
be a noticeable effect
in the case of J0007,
as the particle cooling time,
$\tau_c=100(1+0.144B^2_{-6})(E_\gamma/1\,\text{TeV})^{-0.5}$ kyr, is equal to the pulsar spin-down age for a magnetic field of 2.5 $\mu$G, which is comparable to the field found in the SED modeling.
The model SED amplitude is
$A=(1.88^{+0.13}_{-1.21})\times10^{34}$ particles eV$^{-1}$,
the break energy is $E_\text{b}=1.8^{+12}_{-0.4}$ TeV,
and the slopes before and after the break are 
$p_1=1.68^{+0.43}_{-0.03}$ and $p_2=2.68$,
which are very close to the values measured from the fits to the 
un-cooled ($p_1=1.68$ for $\Gamma=1.34$)
and cooled ($p_2=2.7$ for $\Gamma=1.85$)
CXO spectra, respectively.

The 
best-fit
magnetic field 
$B=1.41^{+0.05}_{-0.04}~\mu$G,
is lower than the typically assumed  ISM value,
$B_\textbf{ISM}\approx3-5~\mu$G, 
which is commonly seen when modeling the broadband SED of PWNe
(see e.g. \citealt{2024ApJ...968...67G}, \citealt{2022ApJ...930..148B}.) 
It may be 
caused by the rapid expansion of the pulsar wind within the SNR,
creating a wind-blown bubble.
It can be seen in the fit shown in Fig. \ref{fig:sed} that this simple model struggles to simultaneously fit the X-ray and 
$\gamma$-ray data. 
The onset of the exponential decay
(governed by $E_c$)
begins too early in the synchrotron model component, 
resulting in a poor match to the observed slope of the X-ray spectrum, while the IC component provides a good fit to the $\gamma$-ray data.

We also modeled the broadband SED and angular sizes of the CTA~1 system using the one-zone
dynamical evolution model of \citet{2009ApJ...703.2051G}.
This model self-consistently tracks the time evolution of both the SNR and the PWN,
predicting the synchrotron and inverse Compton (IC) emission spectra, as well as
the angular sizes of both components.
The evolutionary dynamics are governed by twelve free parameters describing the
supernova remnant, the surrounding medium, the pulsar wind nebula, and the particle injection spectrum.
The pulsar spin-down power $\dot{E}$ and characteristic age $\tau_{\rm c}$,
which are directly measured from the pulsar timing solution \citep{2008Sci...322.1218A},
are used to fix the present-day pulsar properties; the initial spin-down behavior
is parametrized by the spin-down timescale $\tau_{\rm sd}$ and braking index $n$,
both of which are free parameters.

The observational constraints include the SNR and PWN angular sizes, $\theta_{\rm snr} \approx 45 \pm 10'$ and
$\theta_{\rm pwn} \in [10.2, 13.8]'$, are taken from \citet{1997ApJ...485..221S} and \citet{2025SCPMA..6879503L}, respectively.
The radio constraint is a 1.4~GHz flux density upper limit from VLA observations \citep{2013ICRC...33.2656G}.
Gaussian priors are placed on the distance ($d = 1.4 \pm 0.4$~kpc).

We explored the model parameter space using the affine-invariant ensemble
sampler {\tt emcee} \citep{2013PASP..125..306F}.
We ran 96 walkers for 5000 steps with the 'stretch move' algorithm. 
The log-likelihood is evaluated as $\ln \mathcal{L} = -\chi^2 / 2$,
where $\chi^2$ is computed over the angular sizes, radio upper limit, CXO spectral
data, Fermi SED points, and LHAASO WCDA and KM2A fluxes \citep{2025SCPMA..6879503L}.
We present two models that differ primarily in their dynamical history. 
Model A fits the individual CXO spectral points and evolves the PWN through the reverse shock encounter into a post-compression state. Model B instead uses the CXO 0.5--7~keV integrated band flux as a lower limit and the ASCA 0.5--7~keV flux as an upper limit, allowing the pre-compression solution to be explored; the resulting fit places  the PWN in free expansion phase.
Both models converged on a broken power law injection spectrum with $p_1\approx0.7$, $p_2 \approx 2.7 - 2.9$ and $B \approx 1.4 -3.2~\mu$~G.

\begin{figure*}
    \centering
\includegraphics[width=1\linewidth]{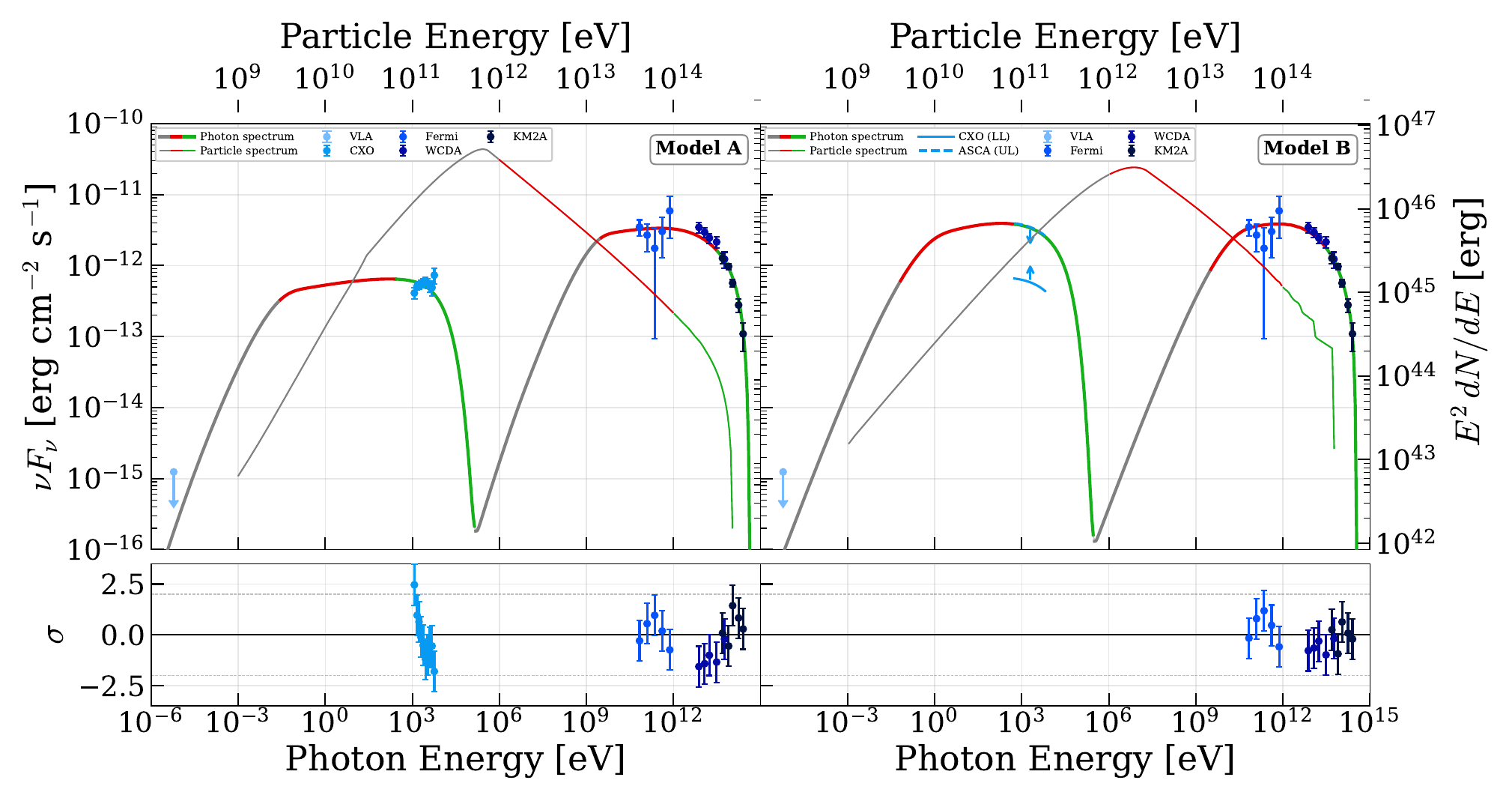}
    \caption{Broadband spectral energy distribution of the J0007 PWN. 
    Data include the VLA upper limit (UL) \citep{2013ICRC...33.2656G}, CXO data (this work), Fermi-LAT SED points (this work), LHAASO WCDA and KM2A fluxes \citep{2025SCPMA..6879503L}. 
    The CXO 0.5--7~keV lower limit and ASCA 0.5--7~keV upper limit used to constrain Model B are shown as blue curves with arrows indicating the bound direction.
    The thick (thin) curve indicates the photon (particle) spectrum of the best-fit one-zone dynamical evolution model \citep{2009ApJ...703.2051G} for Model A (left, post-compression) and Model B (right, pre-compression), with corresponding parameters listed in Table~\ref{tab:model_pars_cta1}. 
    The photon spectra are colored so that the corresponding colors in the particle spectra indicate the particles contributing most of the photon flux at a given energy. 
    The CXO 0.5--7~keV lower limit and ASCA 0.5--7~keV upper limit used to constrain Model B are shown as blue curves with arrows indicating the bound direction.
    Bottom panels show the model residuals.
    }
    \label{fig:sed_pwnmodel}
\end{figure*}

The best-fit model parameters are listed in Table~\ref{tab:model_pars_cta1},
and the predicted broadband emission is compared to the observed fluxes in
Table~\ref{tab:model_obs_cta1} and Figure~\ref{fig:sed_pwnmodel}.
Model A achieves $\chi^2 = 26.4$ for 16 degrees of freedom (reduced $\chi^2 = 1.65$),
providing a good description of the radio upper limit, CXO spectral data,
Fermi-LAT SED points, and LHAASO KM2A fluxes.
The best-fit age of 15.5~kyr is consistent with independent SNR age estimates
of 5--15~kyr \citep{1993AJ....105.1060P,1997ApJ...485..221S,2004ApJ...601.1045S}.
The best-fit magnetic field of $B \approx 1.4~\mu$G is lower than the $B \approx 6~\mu$G
found by \citet{2013ApJ...764...38A} using ASCA X-ray data.
This low field can be attributed to the rapid adiabatic expansion of the nebula, and the discrepancy with \citet{2013ApJ...764...38A} can be attributed to the larger ASCA flux, 
which may be more indicative of the total X-ray  flux within the TeV PWN extent as ASCA's FOV is able to capture a larger flux region (r = 9$'$) than CXO (see left panel of Figure~\ref{fig:cxo_reg}).
Model A places the reverse shock encounter at $\approx$2.7~kyr, after which the PWN was compressed and is currently re-expanding, placing it in a more evolved dynamical state than assumed in \citet{2013ApJ...764...38A}.
The particle injection spectrum requires a broken power law with a hard low-energy
index $p_1 \approx 0.68$ below $E_{\rm break} \approx 740$~GeV and a steep index
$p_2 \approx 2.85$ above the break, consistent with the extended nebula photon index
$\Gamma = 1.85 \pm 0.11$ measured in Section~\ref{sec:spectral} via
$p = 2\Gamma - 1 \approx 2.7$ (see Table~\ref{tab:fit_stats}).
Model B is also a good fit to the data  ($\chi^2/\text{dof} = 6.96/8$) with $p_1 \approx 0.72$, $p_2 \approx 2.68$, and a PWN age of $\approx 9$~kyr, corresponding to a pre-compression evolutionary phase. In this case the best-fit pushes the  model's X-ray flux to the upper limit set by ASCA flux which, in turn, increases magnetic field to $B \approx 3.2~\mu$G. 

We also tested the possibility that the PWN injection spectrum follows a power law with $p_2 = 1.6$, as suggested by the X-ray photon index of the torus. 
In this scenario, we found no combination of dynamical parameters that could simultaneously reproduce the Fermi-LAT and LHAASO fluxes using the CMB as the sole IC target photon field, with the Fermi fluxes systematically under-predicted by the best-fit models.
An acceptable fit (reduced $\chi^2_\nu \approx 2$) was obtained with an additional starlight ($T \approx 3000$~K) photon field, with energy density $u \gtrsim 1000~{\rm eV}~{\rm cm}^{-3}$, requiring proximity to a dense stellar cluster.

\begin{table*}
    \caption{PWN Model Input Parameters}
    \label{tab:model_pars_cta1}
    \centering
    \begin{tabular}{lcc}
    \hline
    \hline
    {\sc Parameter} & {\sc Model A} & {\sc Model B} \\
    \hline
    \multicolumn{3}{c}{\it SNR Parameters} \\
    \hline
    Explosion Energy $E_{\rm sn}$ & $1.1\times10^{51}$~erg & $1.1\times10^{51}$~erg \\
    Ejecta Mass $M_{\rm ej}$ & 10.5~$M_\odot$ & 13.4~$M_\odot$ \\
    ISM Density $n_{\rm ism}$ & 1.55~cm$^{-3}$ & 0.068~cm$^{-3}$ \\
    Distance $D$ & 1.26~kpc & 1.45~kpc \\
    \hline
    \multicolumn{3}{c}{\it Pulsar Parameters} \\
    \hline
    Spin-down Timescale $\tau_{\rm sd}$ & 137~yr & 4710~yr \\
    Braking Index $n$ & 2.78 & 2.96 \\
    \hline
    \multicolumn{3}{c}{\it PWN Parameters} \\
    \hline
    Wind Magnetization $\eta_{\rm B}$ & $1.27\times10^{-4}$ & $9.84\times10^{-2}$ \\
    Minimum e$^{\pm}$ Injection Energy $E_{\rm min}$ & 30~GeV & 1.14~GeV \\
    Maximum e$^{\pm}$ Injection Energy $E_{\rm max}$ & 470~TeV & 379~TeV \\
    Break Energy $E_{\rm break}$ & 740~GeV & 2622~GeV \\
    Particle Index $p_1$ & 0.68 & 0.72 \\
    Particle Index $p_2$ & 2.85 & 2.68 \\
    \hline
    \hline
    \end{tabular}
\end{table*}

\begin{table*}[ht!]
\caption{Observed properties of the CTA~1 system, alongside the model predicted properties}
    \label{tab:model_obs_cta1}
    \resizebox{0.95\linewidth}{!}{
    \centering
    \begin{tabular}{lcccc}
    \hline
    \hline
    {\sc Property} & {\sc Observed} & {\sc Model A} & {\sc Model B} & {\sc References} \\
    \hline
    \multicolumn{5}{c}{\it Pulsar Properties} \\
    \hline
    $\dot{E}$ & $4.5\times10^{35}$~erg~s$^{-1}$ & Fixed & Fixed & 1 \\
    $\tau_{\rm char}$ & 13.9~kyr & Fixed & Fixed & 1 \\
    \hline
    \multicolumn{5}{c}{\it Pulsar Wind Nebula Properties} \\
    \hline
    SNR Radius $\theta_{\rm snr}$ [arcmin] & $45\pm10$ & 44.4 & 52.8 & 2 \\
    PWN Radius $\theta_{\rm pwn}$ [arcmin] & $[10.2,\,13.8]$ & 11.4 & 14.5 & 3 \\
    1.4~GHz Flux Density$^a$ & $<1.24\times10^{-15}$ & $1.78\times10^{-16}$ & $8.83\times10^{-17}$ & 4 \\
    1.18~keV$^b$ & $(4.11\pm0.73)\times10^{-13}$ & $5.92\times10^{-13}$ & --- & This work \\
    1.46~keV$^b$ & $(5.16\pm0.62)\times10^{-13}$ & $5.76\times10^{-13}$ & --- & This work \\
    1.69~keV$^b$ & $(5.24\pm0.61)\times10^{-13}$ & $5.63\times10^{-13}$ & --- & This work \\
    1.98~keV$^b$ & $(5.56\pm0.66)\times10^{-13}$ & $5.49\times10^{-13}$ & --- & This work \\
    2.37~keV$^b$ & $(5.73\pm0.83)\times10^{-13}$ & $5.29\times10^{-13}$ & --- & This work \\
    2.83~keV$^b$ & $(6.07\pm0.83)\times10^{-13}$ & $5.07\times10^{-13}$ & --- & This work \\
    3.36~keV$^b$ & $(5.65\pm0.83)\times10^{-13}$ & $4.83\times10^{-13}$ & --- & This work \\
    3.99~keV$^b$ & $(5.13\pm0.93)\times10^{-13}$ & $4.57\times10^{-13}$ & --- & This work \\
    4.77~keV$^b$ & $(4.90\pm1.14)\times10^{-13}$ & $4.27\times10^{-13}$ & --- & This work \\
    5.71~keV$^b$ & $(7.32\pm1.87)\times10^{-13}$ & $3.94\times10^{-13}$ & --- & This work \\
    CXO 0.5--7~keV$^c$ [LL] & $>1.4\times10^{-12}$ & --- & $8.13\times10^{-12}$ & This work \\
    ASCA 0.5--7~keV$^c$ [UL] & $<8.4\times10^{-12}$ & --- & $8.13\times10^{-12}$ & 5 \\
    67~GeV$^b$ & $(3.50\pm0.90)\times10^{-12}$ & $3.23\times10^{-12}$ & $3.35\times10^{-12}$ & This work \\
    123~GeV$^b$ & $(2.69\pm1.13)\times10^{-12}$ & $3.30\times10^{-12}$ & $3.58\times10^{-12}$ & This work \\
    224~GeV$^b$ & $(1.75\pm1.66)\times10^{-12}$ & $3.34\times10^{-12}$ & $3.72\times10^{-12}$ & This work \\
    407~GeV$^b$ & $(3.02\pm1.73)\times10^{-12}$ & $3.35\times10^{-12}$ & $3.82\times10^{-12}$ & This work \\
    741~GeV$^b$ & $(5.89\pm3.48)\times10^{-12}$ & $3.31\times10^{-12}$ & $3.84\times10^{-12}$ & This work \\
    7.5~TeV$^b$ & $3.46^{+0.56}_{-0.59}\times10^{-12}$ & $2.56\times10^{-12}$ & $3.01\times10^{-12}$ & 3 \\
    11.8~TeV$^b$ & $2.95^{+0.47}_{-0.45}\times10^{-12}$ & $2.30\times10^{-12}$ & $2.65\times10^{-12}$ & 3 \\
    17.2~TeV$^b$ & $2.44^{+0.36}_{-0.42}\times10^{-12}$ & $2.05\times10^{-12}$ & $2.32\times10^{-12}$ & 3 \\
    30.5~TeV$^b$ & $2.15^{+0.35}_{-0.41}\times10^{-12}$ & $1.64\times10^{-12}$ & $1.77\times10^{-12}$ & 3 \\
    57.3~TeV$^b$ & $1.23^{+0.30}_{-0.33}\times10^{-12}$ & $1.16\times10^{-12}$ & $1.17\times10^{-12}$ & 3 \\
    49~TeV$^b$ & $1.27^{+0.20}_{-0.22}\times10^{-12}$ & $1.29\times10^{-12}$ & $1.32\times10^{-12}$ & 3 \\
    79~TeV$^b$ & $9.72^{+1.02}_{-0.98}\times10^{-13}$ & $9.17\times10^{-13}$ & $8.78\times10^{-13}$ & 3 \\
    109~TeV$^b$ & $5.67^{+0.67}_{-0.77}\times10^{-13}$ & $6.70\times10^{-13}$ & $6.13\times10^{-13}$ & 3 \\
    175~TeV$^b$ & $2.78^{+0.52}_{-0.69}\times10^{-13}$ & $3.28\times10^{-13}$ & $2.83\times10^{-13}$ & 3 \\
    250~TeV$^b$ & $1.09^{+0.39}_{-0.56}\times10^{-13}$ & $1.23\times10^{-13}$ & $9.89\times10^{-14}$ & 3 \\
    \hline
    \hline
    $\chi^2$ (d.o.f.) & & 26.4~(16) & 6.96~(8) & \\
    \hline
    \end{tabular}
    }
\tablenotetext{a}{Flux density in units of erg~s$^{-1}$~cm$^{-2}$~Hz$^{-1}$.}
\tablenotetext{b}{Spectral energy flux $E^2\,dN/dE$ in units of erg~s$^{-1}$~cm$^{-2}$.}
\tablenotetext{c}{Band flux in units of erg~s$^{-1}$~cm$^{-2}$.}
\tablerefs{(1)~\citet{2008Sci...322.1218A}; 
(2)~\citet{2025JApA...46...14G};
(3)~\citet{2025SCPMA..6879503L}; 
(4)~\citet{2013ICRC...33.2656G}; 
(5)~\citet{1997ApJ...485..221S}
}
\end{table*}

\subsection{Physical Properties of the PWN}

\begin{figure}
    \centering
\includegraphics[width=1.0\linewidth]{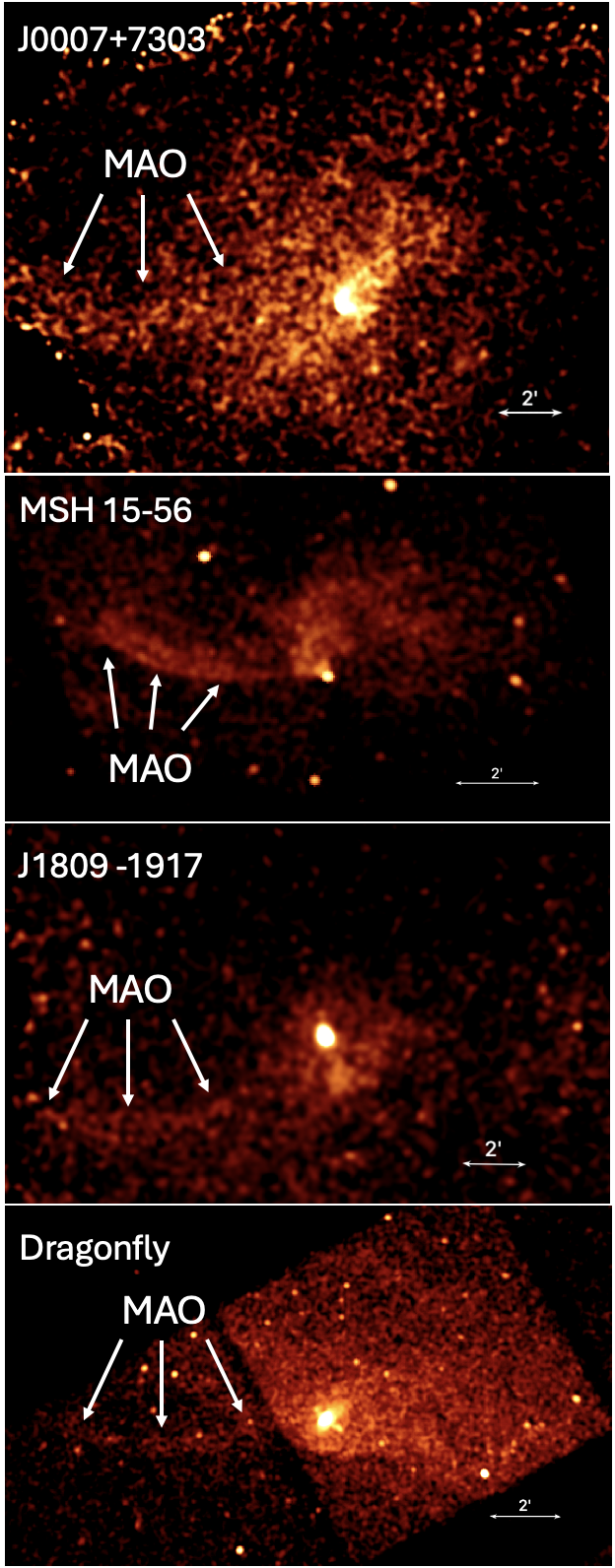}
    \caption{Trans-sonic PWNe with possible faint PXF's}
    \label{fig:mao}
\end{figure}

One can estimate the lower limit of
the magnetic field at the termination shock (TS) 
by assuming that the gyration radius must be smaller than the TS
if particle acceleration happens within the TS (a
commonly considered scenario; see e.g. 
\citealt{2003MNRAS.345..153L},
\citealt{2007A&A...473..683P},
\citealt{2011ApJ...741...39S}),
$B_{\rm TS}\gtrsim70( E_e/500~{\rm TeV})(R_{\rm TS}/2.5\times10^{16}{\rm cm})^{-1}\mu$G.
On the other hand, an upper limit on the magnetic field can be obtained by assuming that only a fraction, $\eta_B$, of injected energy, $\dot{E}$, is in the form of magnetic field
\citep{deOnaWilhelmi2022PeVatrons}, 
$B_{\rm TS}\lesssim200\eta_{\rm B}^{0.5}(R_{\rm TS}/2.5\times10^{16}{\rm cm})^{-1}(\dot{E}/4.5\times10^{35}{\rm erg \,s^{-1}})^{0.5}\mu$G.
Furthermore, 
by comparing these two expressions for the magnetic field
we can constrain 
$\eta_{\rm B}
\gtrsim0.12$.

An upper limit for
the maximum particle energy follows from equating the 
energy loss
due to the accelerating electric potential
within the TS
to the radiative (synchrotron) losses,
$E_{\rm e,max}\lesssim6.3\eta_e^{0.5}(B_{\rm TS}/{\rm 100 \mu G})^{-0.5}$ PeV,
where $\eta_e$ is the ratio of the 
electric field
to the magnetic field
($\eta_e\le1$ in ideal MHD).
The corresponding energy
of photons produced via IC scattering of the CMB
is estimated as
$E_{\rm \gamma,max}\lesssim1.1\eta_e^{0.65}(B_{\rm TS}/{\rm 100 \mu G})^{-0.65}$ PeV.
This 
is compatible with the observed maximum photon energy for the $B_{TS}$ range estimated above.

The above estimates of particle energies  likely also apply to particles producing synchrotron emission in the jets if the jets are formed as a result of backflow from the post-TS region
(see \citealt{2005MNRAS.358..705B}).  However, we do not know if the UHE-emitting particles can be confined within the jets because the jets are not resolved at TeV energies. 
The maximum energy of particles that we can confirm are in the jet
must then be estimated from the energies of synchrotron X-ray photons detected in the CXO images
and the magnetic field constraints obtained above.
The particles producing X-ray synchrotron emission in the 1-7 keV range 
have maximum energies of ${E_e}=7({\rm E_X}/{\rm7\,keV})^{0.5}({\rm B}/{\rm 70\,\mu G})^{-0.5}$ TeV.
The corresponding gyro-radius of these particles is $\sim10^{14}$cm, which is well within the 
radius of the jet $\sim10^{16}$cm (at d=1.4 kpc).
The gyro-radius becomes comparable to the jet radius for particles
with energies
of a few hundred TeV.
Therefore, we can confirm that 
the X-ray-emitting
particles 
can be easily confined
within the jet 
and radiate along its extent.

As the jet propagates it exhibits a bend at a distance R$_b$ from the pulsar, caused by interaction with the external medium.
We
can use this and other dynamical properties of the jet to estimate
the number density of the ambient medium as
\begin{align}
n &= 0.14
\left(\frac{\xi_j}{0.1}\right)
\left(\frac{v_j}{0.5c}\right)
\left(\frac{\dot{E}}{4.5\times10^{35}\,{\rm erg\,s^{-1}}}\right) \notag \\
&\quad \times
\left(\frac{R_b}{2\times10^{17}\,{\rm cm}}\right)^{-1}
\left(\frac{R_j}{6\times10^{16}\,{\rm cm}}\right)^{-1} \notag \\
&\quad \times
\left(\frac{v_{\rm psr}}{100\,{\rm km\,s^{-1}}}\right)^{-2},
\; {\rm cm^{-3}} 
\end{align}
following \citep{2017ApJ...835...66P},
where $\xi_j$ is the energy flow down the jet as a fraction of $\dot{E}$, $v_j$ is the flow velocity of particles in the jet, ${ R_b}$ is the 
distance from the pulsar at which the jet bends
due to interaction with the external medium, ${R_j}$ is the cross-sectional radius of the jet, and $v_{\rm psr}$ is the velocity of the pulsar.

The combined ACIS-I image also reveals an elongated 
structure
among the diffuse emission
extending $\approx$ 8$'$ to the East of the pulsar,
resembling a faint
pulsar X-ray filament (PXF)
which so far has been associated with pulsars  that escaped their SNRs and are moving in the cold ISM with a low sound speed
\citep{2024ApJ...976....4D}.
Due to poor statistics it is not possible to test if the
structure maintains a hard spectrum along its extent as 
``conventional''  PXF's do.
There are several cases where the elongated and slightly bent features are seen in the CXO images of other transonic PWNe with ages comparable to  that of J0007 (see Fig. \ref{fig:mao}).
At least some of these PWNe are still within their host SNRs and curvature could result from the motion of the surrounding SNR medium into which wind particles escaped.

As the pulsar wind continues propagating outwards on larger spatial scales of a few to tens of arcminutes,
for diffuse, roughly symmetrical emission 
(as opposed to the tails present in bow-shock PWNe)
the particle transport should become  diffusion dominated. 
The angular extent of 
the region occupied by particles transported
via Bohm diffusion
over their radiative cooling time
can be estimated as
\begin{align}
\delta_{\rm B} =& \left(6 D t_c\right)^{0.5} d^{-1} \notag \\
\approx&60'
\left(\frac{E_e}{125\,{\rm TeV}}\right)^{0.5}
\left(\frac{t_c}{8\,{\rm kyr}}\right)^{0.5} \notag \\
&\times 
\left(\frac{d}{1.4\,{\rm kpc}}\right)^{-1}
\left(\frac{B}{1.5\,\mu{\rm G}}\right)^{-0.5} ,
\end{align}
where the particle energy corresponds to a more typical
(rather than the maximum energy)
particle
which would radiate ${E_X}=7({E_e}/{\rm 125\,TeV})^2(B/{\rm 1.5\,\mu G})$ keV synchrotron photons 
and 
$E_\gamma = 25(E_e/{\rm 125\,TeV})^{1.3}$ TeV
IC photons.
This angular extent is a factor of a few larger than the observed extent in X-rays
and UHE $\gamma$-rays ($\sim10'$), 
suggesting that particles are diffusing slower than the Bohm mechanism.
This is another common finding seen in many PWNe
associated with TeV halos
\citep{2024NCimR..47..399A}.
If we assume a more realistic 
diffusion coefficient, $D(E_e)=D_0(E_e/E_{e,0})^{\delta}$, with $D_0=5\times 10^{28}$ cm$^2$ s$^{-1}$ at $E_{e,0}=4$ GeV and $\delta=0.4-0.5$ inferred for local ISM  \citep{2017PhRvD..95h3007Y}, 
we obtain an even larger diffusion size (radius) of tens of degrees.  

\section{Summary}

We have presented a detailed analysis of new and archival \textit{Chandra} observations of the PWN powered by PSR~J0007+7303 in CTA~1, supplemented by \textit{Fermi}-LAT data analysis and broadband SED modeling. The combined X-ray dataset resolves the compact nebular morphology, revealing a $\sim20^{\prime\prime}$ jet extending south of the pulsar and bending toward the southwest, a faint counter-jet to the north, and a compact torus oriented approximately perpendicular to the jet axis. The observed jet bending is likely influenced by interaction with the ambient medium or the SNR reverse shock. Relative astrometry over a $\sim20$ yr baseline constrains the pulsar’s transverse velocity to $\lesssim 200~\mathrm{km~s^{-1}}$ (at 95\% confidence for $d=1.4$ kpc), significantly lower than estimates based on its displacement from the 
SNR center, suggesting either an older system age or asymmetric SNR expansion.

Spatially resolved spectroscopy reveals
hard spectra in the compact PWN components, with photon indices $\Gamma \approx 1.2$--1.4 for the jet and torus, indicating minimal synchrotron cooling and suggesting that these regions probe the intrinsic particle acceleration spectrum. In contrast, the extended nebula exhibits a softer spectrum with $\Gamma = 1.85 \pm 0.11$, consistent with radiative cooling. Modeling of the torus morphology yields a viewing angle $\zeta \approx 50^\circ$, consistent with constraints from $\gamma$-ray pulse profile modeling and implying a magnetic inclination in the range $\alpha \sim 20^\circ$--$70^\circ$. The relative brightness of the torus and jet further supports a geometry with a 
larger inclination angle.

The hard X-ray spectra of the compact PWN, 
combined with the low X-ray radiative efficiency ($\sim 10^{-5}$), 
could be explained by the efficient particle acceleration with a relatively low pair multiplicity. The observed spectral hardness is consistent with scenarios in which particle acceleration is dominated by magnetic reconnection in the striped pulsar wind. Broadband SED modeling using a one-zone leptonic scenario
provides a reasonable description of the system
that yields a low magnetic field $B \approx 1.4$--$3.2~\mu\mathrm{G}$ and a high electron cutoff energy $E_{\rm cut} \sim 0.2$--$0.3~\mathrm{PeV}$. Time-dependent dynamical modeling yields similar magnetic field strengths and indicates that the system may be either in a post-reverse-shock compression phase or still in free expansion, depending on the adopted X-ray flux from the TeV-emitting volume.

Overall, CTA~1 emerges as a low-magnetization, low-efficiency PWN capable of accelerating particles to PeV energies, providing an important laboratory for studying particle acceleration and evolution in young pulsar wind nebulae.

\begin{acknowledgements}
The authors are grateful to Svanik Tandon and Reshmi Mukherjee for their insights related to the treatment of the previous VERITAS analysis. Support for this work was provided by the National Aeronautics and Space Administration through Chandra Award Number GO3-24057X issued by the Chandra X-ray Center, which is operated by the Smithsonian Astrophysical Observatory 
under contract NAS8-03060. 

This paper employs a list of Chandra datasets, obtained by the Chandra X-ray Observatory, contained in the Chandra Data Archive (CDA) 

\end{acknowledgements}

\software{CIAO v4.15 \citep{2006SPIE.6270E..1VF}, Naima \citep{2015ICRC...34..922Z}, Sherpa \citep{2001SPIE.4477...76F}, Wavdetect \citep{Freeman2002} }

\bibliographystyle{aasjournal}
\bibliography{references.bib}

\appendix

\section{Astrometry
\label{sec:astrometry}}

The X-ray sources are detected in each observation using the CIAO tool \texttt{wavdetect}.
Prior to running \texttt{wavdetect} the images were binned up by a factor of 2
(pixel size $\approx 1''$)
and restricted to 0.7-8 keV. 
A 12$'\times10'$ 
region centered on (R.A., Decl.)=(0$^{\rm h}$ 06$^{\rm m}$ 48$^{\rm s}$.8352, +73$^\circ$ 02' $38.227''$),
at an angle of 338.88$^\circ$ 
to align with the ACIS-S chips
and excluding the compact PWN region,
was used
to trim the images.
Each of the epoch 2 observations are then aligned to the epoch 1 observation (ObsID 3835).
The CIAO tool \texttt{wcs\_match} is used to determine 
the translational shift minimizing the sum of positional offsets between the matched source pairs
using a radius of 2$''$
and a residual limit of 1
to match sources between observations.
We update the WCS information and the aspect history files accordingly using \texttt{wcs\_update}.

The 1$\sigma$ uncertainty of the RA and Decl. components of the astrometric correction is calculated
for each observation's
alignment to the reference 
using the weighted root-mean-square residual (WRMSR) as e.g.
\begin{equation}
    {\rm WRMSR_{RA}^2} = \frac{\sum_{i=1}^N w_{{\rm RA},i} R_{{\rm RA},i}^2}{\sum_{i=1}^N w_{{\rm RA},i}}
\end{equation}

with 

\begin{equation}
    w_{{\rm RA},i} = \frac{1}{\sigma_{{\rm RA},i}^2} = \frac{1}{\sigma_{{\rm ref},i}^2+\sigma_{{\rm X},i}^2}
\end{equation}

where $N$ is the number of matching sources found between each pair of observations
after the final transformation is computed by 
\texttt{wcs\_match}, $R_{{\rm RA},i}$ is the residual in RA for the $i$th pair after the astrometric correction, and $\sigma_{{\rm ref},i}$ and $\sigma_{{\rm X},i}$ are the RA positional uncertainties of the reference source and the X-ray source to be aligned with of the $i$th pair.
The solutions are summarized in Table \ref{tab:astro_solution}.
Corrected (with \texttt{wcs\_update}) images from  epoch 2 are co-added to create a deeper merged image using the CIAO task \texttt{merge\_obs}.
To determine the RA and Decl. components of the astrometric error for the merged image a weighted average of the WRMSR's is computed
according to e.g.

\begin{equation}
    \sigma_{\rm RA, sys}^2 = \frac{\sum_{i=1}^N t_i {\rm WRMSR}_{{\rm RA},i}^2}{\sum_{i=1}^N t_i}
\end{equation}

where $N$ is the number of observations and $t_i$ is the exposure time of each observation. The Decl. component
is computed similarly,
and the errors were found to be 
$\sigma_{\rm RA, sys} = 11$ mas yr$^{-1}$
and 
$\sigma_{\rm Decl, sys} = 14$ mas yr$^{-1}$.

\begin{deluxetable*}{lcccccr}[b]
\tablecaption{Astrometry Solutions\label{tab:astro_solution}}
\tablewidth{0pt}
\tablehead{
\colhead{ObsID} &
\colhead{$\Delta$R.A.($\arcsec$)} &
\colhead{$\Delta$Decl.($\arcsec$)} &
\colhead{${\rm WRMSR_{RA}}$($\arcsec$)} &
\colhead{${\rm WRMSR_{Decl}}$($\arcsec$)} &
\colhead{\# of pairs}
}
\startdata
26662 & 0.81 & 0.17 & 0.18 & 0.15 & 6 \\
27102 & -1.14 & 0.92 & 0.26 & 0.40 & 6 \\
27103 & 0.62 & -0.40 & 0.25 & 0.24 & 7 \\
27104 & -1.36 & 1.57 & 0.14 & 0.21 & 4 \\
27105 & -0.37 &  -0.51 & 0.26 & 0.39 & 7 \\ 
29118 & -0.74 & 1.34 & 0.19 & 0.15 & 7 \\
\enddata
\tablecomments{
Astrometry solutions for the later 6 observations with the reference ObsID=3835.  
A source match radius of 2$''$
and a residual limit of 1$''$ were used for \texttt{wcs\_match} in each observation. 
Shown are the change in pointing RA and Decl. of each observation after alignment, 
the RA and Decl. components of the
WRMSR
computed over all source pairs in each observation,
and the number of source pairs used for alignment.
}
\label{tab:astrometry}
\end{deluxetable*}

\end{document}